\documentclass[conference]{IEEEtran}
\pdfoutput=1
\IEEEoverridecommandlockouts

\usepackage{cite}
\usepackage{amsmath,amssymb,amsfonts}
\usepackage{algorithmic}
\usepackage{graphicx}
\usepackage[acronym]{glossaries}  
\usepackage{multicol}
\usepackage{multirow}
\usepackage{xcolor}
\usepackage{booktabs}
\usepackage{comment}
\def\BibTeX{{\rm B\kern-.05em{\sc i\kern-.025em b}\kern-.08em
    T\kern-.1667em\lower.7ex\hbox{E}\kern-.125emX}}

\begin{document}

\title{{\fontsize{22}{12}\selectfont End-to-End Efficiency in Keyword Spotting: A System-Level Approach for Embedded Microcontrollers}}


\author{
\IEEEauthorblockN{
Pietro Bartoli\IEEEauthorrefmark{1}, 
Tommaso Bondini\IEEEauthorrefmark{1},
Christian Veronesi\IEEEauthorrefmark{1}, 
Andrea Giudici\IEEEauthorrefmark{1},
Niccolò Antonello\IEEEauthorrefmark{2},
Franco Zappa\IEEEauthorrefmark{1}
}
\IEEEauthorblockA{\textit{\IEEEauthorrefmark{1}Dept. of Electronics, Information and Bioengineering, Politecnico di Milano, Milan, Italy}}
\IEEEauthorblockA{\textit{\IEEEauthorrefmark{2}Smart Eyewear Laboratory, EssilorLuxottica, Milan, Italy}
}
  \thanks{© 2025 IEEE. Personal use of this material is permitted. 
  Permission from IEEE must be obtained for all other uses, 
  in any current or future media, including reprinting/republishing 
  this material for advertising or promotional purposes, creating 
  new collective works, for resale or redistribution to servers or 
  lists, or reuse of any copyrighted component of this work in 
  other works.}
}
\maketitle

\begin{abstract}
Keyword spotting (KWS) is a key enabling technology for hands-free interaction in embedded and IoT devices, where stringent memory and energy constraints challenge the deployment of AI-enabeld devices. 
In this work, we systematically evaluate and compare several state-of-the-art lightweight neural network architectures, including DS-CNN, LiCoNet, and TENet, alongside our proposed Typman-KWS (TKWS) architecture built upon MobileNet, specifically designed for efficient KWS on microcontroller units (MCUs). 
Unlike prior studies focused solely on model inference, our analysis encompasses the entire processing pipeline, from Mel-Frequency Cepstral Coefficient (MFCC) feature extraction to neural inference, and is benchmarked across three STM32 platforms (N6, H7, and U5). 
Our results show that TKWS with three residual blocks achieves up to 92.4\% F1-score with only 14.4k parameters, reducing memory footprint without compromising the accuracy. 
Moreover, the N6 MCU with integrated neural acceleration achieves the best energy-delay product (EDP), enabling efficient, low-latency operation even with high-resolution features. 
Our findings highlight the model accuracy alone does not determine real-world effectiveness; rather, optimal keyword spotting deployments require careful consideration of feature extraction parameters and hardware-specific optimization.
\end{abstract}


\section{Introduction}
Keyword spotting (KWS) refers to the detection of predefined words in a continuous audio stream, enabling hands-free interaction with devices and forming a fundamental component of voice-controlled systems and modern IoT applications~\cite{Lopez_2022}. 
Widely deployed consumer platforms like \textit{Amazon Alexa}, \textit{Google Assistant}, and \textit{Apple Siri} exploit KWS to recognize specific wake words (“Alexa,” “Hey Google,” and “Hey Siri”) allowing for intuitive, hands-free voice interfaces~\cite{Michaely_2017}.
In these systems, the KWS module operates locally to reduce latency and maintain privacy, activating cloud-based processing after the wake word is detected for advanced tasks like speech recognition~\cite{Ku_2024,Jabbeen_2024}.
However, while these KWS algorithms run efficiently on consumer devices such as smart speakers and smartphones, with their greater processing power, deploying them on battery-powered microcontrollers (MCUs) introduces significant challenges~\cite{Ulkar_2021}. 
MCUs operate under strict memory and computation constraints, requiring KWS solutions to carefully balance responsiveness and accuracy~\cite{Cioflan_2024,Wang_2022}, while also minimizing energy consumption, a critical factor in battery-powered device.
This trade-off between latency, accuracy, and efficiency becomes even more complex as recent deep learning approaches, although improving accuracy~\cite{Arik_2017}, often exceed the computational and memory budgets of embedded hardware~\cite{Martinez_2025}.


Moreover, the audio pre-processing chain imposes additional overhead, particularly through Mel-Frequency Cepstral Coefficients (MFCCs), one of the most widely adopted feature extraction methods in automatic speech processing~\cite{Vrea_2024}.
MFCCs are acoustic features that mimic human auditory perception by reflecting the ear’s nonlinear frequency sensitivity, achieved first converting audio to a mel-scaled spectrogram and then applying dimensionality reduction via the Discrete Cosine Transform (DCT)~\cite{Abdul_2022}. 

While most previous works have focused on optimizing KWS model architectures~\cite{DSCNN,Blouw_2021,Mittermaier_2019}, there is still a lack of comprehensive studies that evaluate both the energy and memory costs of audio pre-processing and inference on different embedded platforms. 
To fill this gap, in this work we systematically assess the full KWS pipeline, including audio pre-processing, feature extraction and neural network inference on three different MCUs: U5, H7, and N6.

We also introduce \textit{Typman-KWS} (TKWS), a lightweight neural model designed for low memory and energy usage, and we compared it with different model architectures for KWS on MCUs. 
Through experimental analysis, we systematically quantify the trade-offs between latency, accuracy, and resource utilization, providing practical guidelines for the deployment of efficient KWS solutions in energy-constrained embedded systems.

\begin{figure*}[t]
    \centering
    \includegraphics[width=\textwidth]{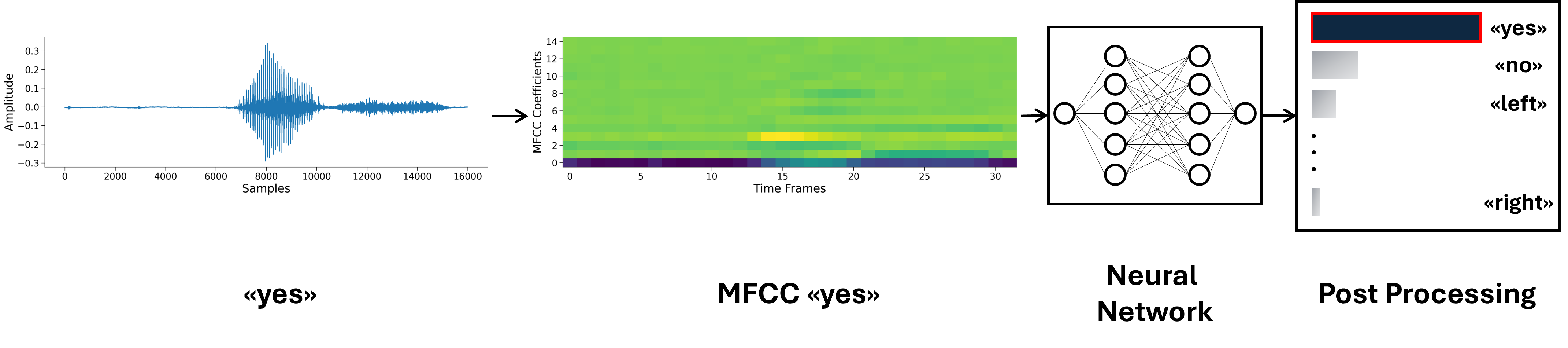}
    \caption{\small Overview of the processing pipeline. The raw audio waveform is transformed into time-frequency features (\textit{MFCC}), which are subsequently processed by a neural network. A post-processing stage is then applied to extract the most likely prediction.}
    \label{fig:pipeline}
\end{figure*}

\section{Experimental Setup}
The experimental setup was carefully designed to achieve two principal aims, focusing on both model validation and performance profiling.
First, we evaluate the performance of the proposed TKWS architecture in terms of accuracy and number of parameters, comparing it to reference models from the literature.
Second, we aim to quantify the energy requirements of deploying a KWS system on MCUs by systematically analyzing how variations in MFCC feature extraction parameters impact the overall resource consumption of the pipeline.

\subsection{Dataset}
All models were trained and evaluated on the \textit{Google Speech Commands Dataset} (GSCD)~\cite{Warden_2018} v0.02, using a subset of 10 keyword classes —\textit{yes}, \textit{no}, \textit{up}, \textit{down}, \textit{left}, \textit{right}, \textit{on}, \textit{off}, \textit{stop}, and \textit{go} — for the classification task.

To improve robustness to real-world conditions, training samples were augmented with noise from the GSCD, including both synthetic and real background sounds, mixed at a signal-to-noise ratio (SNR) sampled from \( \mathcal{N}(10\,\mathrm{dB}, 5\,\mathrm{dB}) \).
The pipeline (Fig.~\ref{fig:pipeline}) extracts and normalizes MFCC features from raw audio before neural classification.
To assess the impact of feature resolution, we evaluated multiple MFCC configurations by varying the number of windows (32, 63) and Mel filter banks (15, 30). 
These values were selected to balance efficiency and expressiveness: 32 and 63 time frames align with SFFT sizes of 1024 and 512, avoiding zero-padding and enabling efficient computation, while 15 and 30 Mel filters represent lightweight and more detailed spectral resolutions.

\subsection{Proposed Models}
We evaluate four lightweight neural network architectures for KWS on low-power embedded platforms, including three reference models from the literature and TKWS, a custom design developed to meet the memory and energy constraints of MCUs.
The evaluated architectures are:

\subsubsection{DS-CNN}
Implements depthwise separable convolutions by factorizing standard convolutions into depthwise and pointwise operations, as in the Hello Edge framework~\cite{DSCNN}, to achieve a favorable balance between accuracy and resource efficiency.

\subsubsection{LiCO-Net}
Uses the small configuration LiCO-Net~\cite{Yang_2022} (LicoNet-S), adapted from its original streaming inference design with causal convolutions to perform one-shot inference by modifying the padding strategy, enabling full sample processing without maintaining internal state.

\subsubsection{TENet}
Employs a multi-branch temporal convolutional architecture available in standard (TENet6) and narrow (TENet6-N) versions~\cite{Li_2020}, utilizing kernel fusion at inference to efficiently aggregate multi-scale temporal features while minimizing runtime complexity.

\subsubsection{TKWS}
Inspired by the inverted bottleneck design of MobileNetV2~\cite{Sandler_2019}, the model is built upon stacked residual blocks that combine pointwise expansion, double 1D depthwise convolutions, and projection layers. 
It operates directly on MFCC temporal sequences and is implemented in two configurations, featuring either 2 (TKWS-2) or 3 (TKWS-3) residual blocks.

For all models, the classification head consists of a global average pooling layer followed by a fully connected layer with softmax activation to output class probabilities.

\begin{figure*}[t]
\centering
\includegraphics[width=\linewidth]{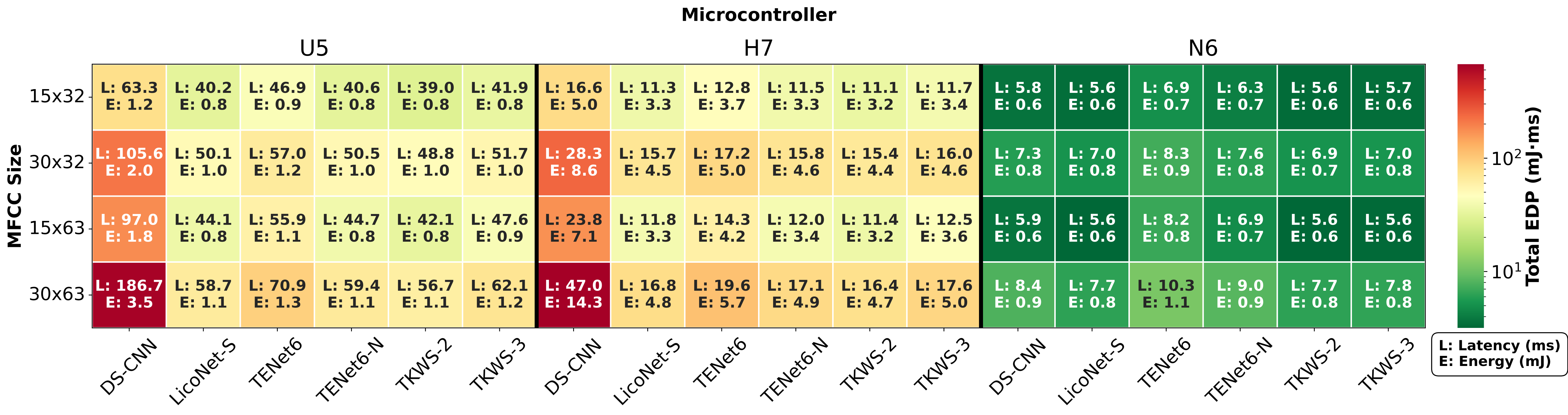}
\caption{\small Heatmap of total Energy-Delay Product (EDP) [\(mJ \cdot ms\)] for the end to end keyword spotting pipeline, evaluated across different MFCC sizes (Mel filter banks\(\times\)\#windows), model architectures, and microcontroller platforms (U5, H7, N6). 
Cell colors reflect EDP values (lower is better), while each cell also reports latency (L) in \(ms\) and energy (E) in \(mJ\).}

\label{fig:edp_heatmap}
\end{figure*}

\subsection{Benchmarking Procedure}
The experimental evaluation was performed on three STM32 MCUs-N6 (Cortex-M55), H7 (Cortex-M7), and U5 (Cortex-M33)-representing different computational classes.
All CPUs were configured to operate at their maximum frequencies: 800\.MHz for N6, 480\.MHz for H7, and 160\.MHz for U5.
All models were quantized to 8-bit integers, and MFCC extraction leveraged each platform’s digital signal processor (DSP) for optimal performance.
The evaluated models were assessed according to the following metrics:
\begin{itemize}

    \item \textbf{Number of Parameters}: quantifies the static memory occupation on MCUs, as model weights, which must be stored in flash memory.
    
    \item \textbf{Weighted F1-score}: used as the primary classification performance metric to account for potential class imbalance in the test set.
    
    \item \textbf{Energy-Delay Product (EDP)}: a combined metric balancing latency and energy consumption, defined as the product of latency and energy consumption. 
    This figure of merit provides a comprehensive evaluation of system-level efficiency~\cite{Frattini_2018, Datta_2022}.

\end{itemize}

The overall measurement procedure follows the methodology described in~\cite{Bartoli_2025}.
Latency and energy were measured for the pre-processing and inference phases, and the EDP was calculated by combining these two stages, excluding the negligible post-processing step.

\section{Results and Discussion}
In this section, we present a comprehensive evaluation, beginning with model performance (accuracy and parameter count) and subsequently assessing system-level efficiency on the three aforementioned MCUs.

\begin{table}[t]
\centering
\caption{Model performance metrics for the keyword spotting task. For each model and MFCC size (Mel filter banks\(\times\) timesteps), the table reports the total number of parameters (P) and the weighted F1 score [\%] on the test set.}
\label{tab:model_f1_scores}
\setlength{\tabcolsep}{5pt}
\begin{tabular}{l|*{4}{cc}}
\toprule
\multirow{2}{*}{\textbf{Model}} 
& \multicolumn{2}{c}{\textbf{15×32}} 
& \multicolumn{2}{c}{\textbf{30×32}} 
& \multicolumn{2}{c}{\textbf{15×63}} 
& \multicolumn{2}{c}{\textbf{30×63}} \\
& P & F1 & P & F1 & P & F1 & P & F1 \\
\midrule
DS-CNN     & 46.5k & 86.0 & 46.5k & 86.7 & 46.5k & 91.2 & 46.5k & 91.0  \\
LicoNet-S  & 17.4k & 90.1 & 18.3k & 90.3 & 17.4k & 93.6 & 18.3k & 92.5  \\
TENet6     & 54.2k & 90.5 & 55.7k & 91.2 & 54.2k & 92.9 & 55.7k & 93.2 \\
TENet6-N   & 17.1k & 90.5 & 17.9k & 90.3 & 17.1k & 91.8 & 17.9k & 91.7 \\
TKWS-2     & \textbf{4.6k}  & 86.5 & 5.0k  & 87.4 & \textbf{4.6k}  & 88.8 & 5.1k  & 87.3 \\
TKWS-3     & 14.4k & 90.7 & 14.9k & 91.1 & 14.4k & 92.4 & 14.9k & 91.2 \\
\bottomrule
\end{tabular}
\end{table}

Table~\ref{tab:model_f1_scores} presents the classification performance and parameter count of the evaluated models.
The number of parameter directly impact static memory usage, making model size a critical constraint due to the limited memory resources available on MCUs.
Notably, configurations with 63 time windows consistently outperform those with 32, showing that higher temporal resolution significantly improves accuracy and must be accounted for in resource-constrained system design.
Among the evaluated models, TKWS-3 achieves a high F1 score of 92.4\% with only 14.4k parameters (MFCC size: $15 \times 63$).
This performance closely matches that of LicoNet-S (93.6\% F1, 17.4k parameters), while requiring 17\% fewer parameters.
Compared to DS-CNN (91.2\% F1, 46.5k parameters) and TENet6 (92.9\% F1, 54.2k parameters), TKWS-3 provides equal or better accuracy while reducing the memory footprint by over threefold. 
Even the most compact configuration of TKWS-2 (88.8\% F1 with only 4.6k parameters) demonstrates competitive performance, highlighting the efficiency of the TKWS architecture in achieving high accuracy with minimal memory requirements.

Figure~\ref{fig:edp_heatmap} shows the EDP for the complete KWS pipeline across different MFCC input configurations and model architectures.
Among the platforms evaluated, N6 demonstrates the highest efficiency, leveraging its advanced DSP capabilities for fast MFCC extraction and an integrated NPU that supports low-power, high-speed inference. 
This combination allows the platform to effectively accommodate higher temporal resolutions, such as MFCC inputs with 63 timesteps, without incurring a significant energy cost.
By comparison, both H7 and U5 exhibit worse (i.e., higher) EDP values than N6 and are similar to each other, though for different reasons: the former achieves lower latency with greater energy consumption, while the latter favors energy efficiency over speed.
The similarity in their EDP values indicates that the optimal platform choice should depend on the target application: H7 is preferable when quick response is critical (e.g., real-time command recognition), whereas U5 is more suitable for ultra-low-power scenarios where latency is less important. 

Moreover, the analysis of the EDP highlights the interplay between model architecture and hardware capabilities. 
On MCUs without NPUs, 1D convolutional models consistently achieve better energy efficiency than 2D ones. 
For instance, DS-CNN, based on 2D convolutions, shows the highest EDP on both H7 and U5 due to its greater computational complexity and poor hardware compatibility.
However, on the N6 platform, this trend shifts, as TENet models show higher EDP due to kernel sizes that exceed the NPU's optimal execution range.
As a result, these operations are offloaded to the CPU, leading to increased overhead from data transfers and less efficient layer execution.
These findings underscore the importance of co-designing neural architectures and hardware platforms, as the optimal balance between accuracy and efficiency is ultimately determined by their combined characteristics.

\section{Conclusion}
This study systematically evaluates KWS models on low-power MCUs, assessing both feature extraction and neural inference under real-world conditions.
Results demonstrate that efficient architectures like TKWS-2 and TKWS-3 can significantly reduce memory requirements with minimal accuracy loss, making them well-suited for always-on applications.
The analysis further reveals that hardware capabilities play a critical role: platforms like N6, with integrated DSP and NPU, achieve higher efficiency and speed, especially with complex input features, while general-purpose MCUs involve trade-offs between energy and latency.
Overall, the findings highlight the need for a holistic co-design strategy that aligns model architecture with hardware characteristics to optimize performance and resource utilization in embedded AI systems.

\bibliographystyle{IEEEtran}  
\bibliography{Bibliography.bib}

\end{document}